\documentclass[preprint,prabib,aps]{revtex4}

\begin{document}
\draft
\date{\today}
\title{Ellipsoidal, Cylindrical, Bipolar and Toroidal Wormholes in 5D Gravity }
\author{Sergiu I. Vacaru \thanks{
e--mail: sergiu$_{-}$vacaru@yahoo.com,\ sergiuvacaru@venus.nipne.ro}}
\address{Physics Department, CSU Fresno, Fresno, CA 93740-8031, USA\\
and \\
Centro Multidisciplinar de Astrofisica - CENTRA, Departamento de Fisica,\\
Instituto Superior Tecnico, Av. Rovisco Pais 1, Lisboa, 1049-001,\\
Portugal}
\author{D. Singleton \thanks{
e--mail: dougs@csufresno.edu}}
\address{Physics Department, CSU Fresno, Fresno, CA 93740-8031, USA}

\begin{abstract}
In this paper we construct and analyze new classes of wormhole and flux
tube-like solutions for the 5D vacuum Einstein equations. These 5D solutions
possess generic local anisotropy which gives rise to a gravitational running
or scaling of the Kaluza-Klein ``electric'' and ``magnetic'' charges of
these solutions. It is also shown that it is possible to self--consistently
construct these anisotropic solutions with various rotational 3D
hypersurface geometries ({\it i.e.} ellipsoidal, cylindrical, bipolar and
toroidal). The local anisotropy of these solutions is handled using the
technique of anholonomic frames with their associated nonlinear connection
structures \cite{vst}. Through the use of the anholonomic frames the metrics
are diagonalized, in contrast to holonomic coordinate frames where the
metrics would have off--diagonal components. In the local isotropic limit
these solutions are shown to be equivalent to spherically symmetric 5D
wormhole and flux tube solutions.

\vspace{1cm} PACS: 04.50.+h

\end{abstract}

\maketitle

\vspace{1cm}

\section{Introduction}

The first solutions describing black holes and wormholes in 4D and higher
dimensional gravity were spherical symmetric solutions with diagonal metrics
\cite{mor}. Later Salam, Strathee and Perracci \cite{sal} showed that
including off--diagonal components in higher dimensional metrics is
equivalent to including gauge fields. They concluded that geometrical gauge
fields could act as sources of exotic matter necessary for the construction
of a wormhole. Refs. \cite{chodos,dzhsin} examined locally isotropic
solutions with off--diagonal metric components for 5D vacuum Einstein
equations. These solutions were similar to spherically symmetric 4D wormhole
or flux tube metrics with ``electric'' and/or ``magnetic'' fields running
along the throat of the wormhole. These ``electromagnetic'' fields arose as
a consequence of the off--diagonal elements of the metric. By varying
certain free parameters of the off--diagonal elements of the 5D metrics it
was possible to change the relative strengths of the fields in the
wormhole's throat, and to change the longitudinal and transverse size of the
wormhole's throat. In \cite{vsbd} the we constructed {\it anisotropic}
wormhole and flux tube solutions, which reduced to the solutions of \cite
{chodos,dzhsin} in the isotropic limit. The anisotropy of these metrics was
handled using the method of anholonomic frames with associated nonlinear
connections, which has been developed by one of the authors (SV) in Refs.
\cite{vst}. It was shown that these anisotropic solutions exhibited a
variation or running of the ``electromagnetic'' parameters as a result of the
angular anisotropies and/or through variations of the extra spatial
dimension.

In this paper we extend the investigation of \cite{vsbd} by applying the
anholonomic frames method to construct anisotropic wormhole and flux tube
solutions to 5D Kaluza--Klein theory which possess a range of different
symmetries (elliptic, cylindrical, bipolar, toroidal). We will discuss the
physical consequences of these solutions, in particular the variation
of the ``electromagnetic'' parameters ({\it e.g.} the ``electric''
and ``magnetic'' charges associated with the solutions). This variation
of the ``electromagnetic'' charges, which here occurs in the context
of a higher dimensional gravity theory, can be likened to the variation or
running of electric charge that occurs when a real electric charge is
placed into some dielectric medium or in a quantum vacuum where quantum
fluctuations produce a scale dependent electric charge. We will
sometimes loosely refer to this gravitational variation of the
``electromagnetic'' parameters of the solutions as the gravitational
running, scaling or renormalization of the charges of the solutions.

\section{Anholonomic Frames and 5D Vacuum Einstein Equations}

In this section we outline the basic formulas for 5D Einstein gravity, and
introduce the method of anholonomic frames. We construct locally anisotropic
metrics which are generalizations of those considered in Ref. \cite{vsbd}.
These 5D metrics have a mixture of holonomic and anholonomic variables, and
are most naturally dealt with using anholonomic frames. Finally we analyze
the physical and mathematical properties of these 5D, locally anisotropic
vacuum solutions.

\subsection{Metric ansatz}

Let us consider a 5D pseudo--Riemannian spacetime of signature $(+,-,-,-,-)$
and denote the local coordinates $u^\alpha
=(x^i,y^a)=(x^1,x^2,x^3,y^4=s,y^5=p),$ -- or more compactly $u=(x,y)$ --
where the Greek indices are split into two subsets $x^i$ (holonomic
coordinates) and $y^a$ (anholonomic coordinates) labeled respectively by
Latin indices $i,j,k,...=1,2,3$ and $a,b,...=4,5$. The local coordinate
bases, $\partial _\alpha =(\partial _i,\partial _a),$ and their duals, $%
d^\alpha =\left( d^i,d^a\right) ,$ are written respectively as
\begin{equation}
\partial _\alpha \equiv \frac \partial {du^\alpha }=(\partial _i=\frac %
\partial {dx^i},\partial _a=\frac \partial {dy^a})  \label{pder}
\end{equation}
and
\begin{equation}
d^\alpha \equiv du^\alpha =(d^i=dx^i,d^a=dy^a).  \label{pdif}
\end{equation}
We can treat an arbitrary coordinate, $x^i$ or $y^a,$ as space-like ($x, y, z
$), time-like ($t$) or as the 5$^{th}$ spatial coordinate ($\chi$). The aim
is then to study anisotropies and anholonomic constraints for various
coordinates.

With respect to the coordinate frame base (\ref{pdif}) the 5D
pseudo-Riemannian metric
\begin{equation}
dS^{2}=g_{\alpha \beta }du^{\alpha }du^{\beta }  \label{metric1}
\end{equation}
with its metric coefficients $g_{\alpha \beta }$ parameterized as
\begin{equation}
\left[
\begin{array}{ccccc}
g_{1}+w_{1}^{\ 2}h_{4}+n_{1}^{\ 2}h_{5} & w_{1}w_{2}h_{4}+n_{1}n_{2}h_{5} &
w_{1}w_{3}h_{4}+n_{1}n_{3}h_{5} & w_{1}h_{4} & n_{1}h_{5} \\
w_{1}w_{2}h_{4}+n_{1}n_{2}h_{5} & g_{2}+w_{2}^{\ 2}h_{4}+n_{2}^{\ 2}h_{5} &
w_{2}w_{3}h_{4}+n_{2}n_{3}h_{5} & w_{2}h_{4} & n_{2}h_{5} \\
w_{1}w_{3}h_{4}+n_{1}n_{3}h_{5} & w_{3}w_{2}h_{4}+n_{2}n_{3}h_{5} &
g_{3}+w_{3}^{\ 2}h_{4}+n_{3}^{\ 2}h_{5} & w_{3}h_{4} & n_{3}h_{5} \\
w_{1}h_{4} & w_{2}h_{4} & w_{3}h_{4} & h_{4} & 0 \\
n_{1}h_{5} & n_{2}h_{5} & n_{3}h_{5} & 0 & h_{5}
\end{array}
\right] ,  \label{ansatz0}
\end{equation}
The ansatz functions of this metric are smooth function of the form:
\begin{eqnarray}
g_{1} &=&1,\qquad g_{2,3}=g_{2,3}(x^{2},x^{3})=\epsilon _{2,3}\exp
[2b_{2,3}(x^{2},x^{3})],  \label{bvar} \\
h_{4,5} &=&h_{4,5}(x^{2},x^{3},s)=\exp [2f_{4,5}(x^{2},x^{3},s)],
\label{qvar} \\
w_{1} &=&w_{1}(x^{2}),\qquad w_{2,3}=w_{2,3}(x^{2},x^{3},s),  \nonumber \\
n_{1} &=&n_{1}(x^{2}),\qquad \;n_{2,3}=n_{2,3}(x^{2},x^{3},s);  \nonumber
\end{eqnarray}
The ansatz functions of the metric are taken to depend on two isotropic
variables $(x^{2},x^{3})$ and on one anisotropic variable, $y^{4}=s$.

Metric (\ref{metric1}) can be greatly simplified into the form
\begin{equation}
\delta S^{2}=g_{ij}\left( x\right) dx^{i}dx^{i}+h_{ab}\left( x,s\right)
\delta y^{a}\delta y^{b},  \label{dmetric}
\end{equation}
with diagonal coefficients
\begin{equation}
g_{ij}=\left[
\begin{array}{lll}
1 & 0 & 0 \\
0 & g_{2} & 0 \\
0 & 0 & g_{3}
\end{array}
\right] \mbox{ and }h_{ab}=\left[
\begin{array}{ll}
h_{4} & 0 \\
0 & h_{5}
\end{array}
\right]   \label{ansatzd}
\end{equation}
if instead of coordinate bases (\ref{pder}) and (\ref{pdif}) one used
anholonomic frames (anisotropic bases)
\begin{equation}
{\delta }_{\alpha }\equiv \frac{\delta }{du^{\alpha }}=(\delta _{i}=\partial
_{i}-N_{i}^{b}(u)\ \partial _{b},\partial _{a}=\frac{\partial }{dy^{a}})
\label{dder}
\end{equation}
and
\begin{equation}
\delta ^{\alpha }\equiv \delta u^{\alpha }=(\delta ^{i}=dx^{i},\delta
^{a}=dy^{a}+N_{k}^{a}(u)\ dx^{k})  \label{ddif}
\end{equation}
where the $N$--coefficients are parametrized as
\[
N_{1}^{4}=w_{1},\qquad N_{2,3}^{4}=w_{2,3}\mbox{ and }N_{1}^{5}=n_{1},\qquad
N_{2,3}^{5}=n_{2,3}
\]
They define an associated nonlinear connection (N--connection) structure.
(see Refs \cite{vst,v}). Here, we shall not emphasize the N--connection
formalism. The anisotropic frames (\ref{dder}) and (\ref{ddif}) are
anholonomic because, in general, they satisfy some anholonomic relations,
\[
\delta _{\alpha }\delta _{\beta }-\delta _{\beta }\delta _{\alpha
}=W_{\alpha \beta }^{\gamma }\delta _{\gamma },
\]
with nontrivial anholonomy coefficients
\begin{eqnarray}
W_{ij}^{k} &=&0,\qquad W_{ai}^{k}=0,\qquad W_{ab}^{k}=W_{ab}^{c}=0,
\label{anholonomy} \\
W_{ij}^{a} &=&-\Omega _{ij}^{a},\qquad W_{bj}^{a}=-\partial
_{b}N_{j}^{a},\qquad W_{ia}^{b}=\partial _{a}N_{j}^{b},  \nonumber
\end{eqnarray}
where
\[
\Omega _{ij}^{a}=\delta _{j}N_{i}^{a}-\delta _{i}N_{j}^{a}.
\]
Conventionally, the N--coefficients decompose spacetime objects ({\it e.g.}
tensors, spinors and connections) into objects with mixed
holonomic--anholonomic characteristics. The holonomic parts of an object are
indicated with indices of type $i,j,k,...$, while the anholonomic parts have
indices of type $a,b,c,...$. Tensors, metrics and linear connections with
coefficients defined with respect to anholonomic frames (\ref{dder}) and (%
\ref{ddif}) are distinguished (d) by N--coefficients into holonomic and
anholonomic subsets and are called d--tensors, d--metrics and d--connections.

\subsection{Einstein equations in holonomic--anholonomic variables}

The main ``trick'' of the anholonomic frames method for integrating
Einstein's equations in general relativity and various (super)string and
higher / lower dimension gravitational theories, is to find the coefficients
$N_j^a$ such that the block matrices $g_{ij}$ and $h_{ab}$ are diagonalized
\cite{vst,v}. This greatly simplifies computations. With respect to such
anholonomic frames the partial derivatives are N--elongated (locally
anisotropic).

Metric (\ref{metric1}) with coefficients (\ref{ansatz0}) (or equivalently,
the d--metric (\ref{dmetric}) with coefficients (\ref{ansatzd})) is assumed
to solve the 5D Einstein equations
\begin{equation}
R_{\alpha \beta }-\frac{1}{2}g_{\alpha \beta }R=\kappa \Upsilon _{\alpha
\beta },  \label{5einstein}
\end{equation}
where $\kappa $ and $\Upsilon _{\alpha \beta }$ are respectively
the coupling constant and the energy--momentum tensor. For most
of the paper we will consider vacuum solutions, $\Upsilon _{\alpha
\beta} = 0$. The nontrivial components of the Ricci tensor (details of the
computations are given in Refs. \cite {vsbd,v}), for the ansatz,
are
\begin{eqnarray}
R_{2}^{2} &=&R_{3}^{3}=-\frac{1}{2g_{2}g_{3}}\left[ g_{3}^{\bullet \bullet }-%
\frac{g_{2}^{\bullet }g_{3}^{\bullet }}{2g_{2}}-\frac{(g_{3}^{\bullet })^{2}%
}{2g_{3}}+g_{2}^{^{\prime \prime }}-\frac{g_{2}^{^{\prime }}g_{3}^{^{\prime
}}}{2g_{3}}-\frac{(g_{2}^{^{\prime }})^{2}}{2g_{2}}\right] ,  \label{ricci1a}
\\
R_{4}^{4} &=&R_{5}^{5}=-\frac{\beta }{2h_{4}h_{5}},  \label{ricci1b} \\
R_{42} &=&-w_{2}\frac{\beta }{2h_{5}}-\frac{\alpha _{2}}{2h_{5}},\qquad
\;R_{43}=-w_{3}\frac{\beta }{2h_{5}}-\frac{\alpha _{3}}{2h_{5}},
\label{ricci1c} \\
R_{52} &=&-\frac{h_{5}}{2h_{4}}\left[ n_{2}^{\ast \ast }+\gamma n_{2}^{\ast }%
\right] ,\qquad \;R_{53}=-\frac{h_{5}}{2h_{4}}\left[ n_{3}^{\ast \ast
}+\gamma n_{3}^{\ast }\right] ,  \label{ricci1d}
\end{eqnarray}
where
\begin{eqnarray}
\alpha _{2} &=&{h_{5}^{\ast }}^{\bullet }-\frac{h_{5}^{\ast }}{2}\left(
\frac{h_{4}^{\bullet }}{h_{4}}+\frac{h_{5}^{\bullet }}{h_{5}}\right) ={%
h_{5}^{\ast }}\left( \ln |f_{5}^{\ast }|+f_{5}-f_{4}\right) ^{\bullet
},\qquad {h_{5}^{\ast }\neq 0;}  \label{alpha2} \\
\alpha _{3} &=&{h_{5}^{\ast }}^{\prime }-\frac{h_{5}^{\ast }}{2}\left( \frac{%
h_{4}^{\prime }}{h_{4}}+\frac{h_{5}^{\prime }}{h_{5}}\right) ={h_{5}^{\ast }}%
\left( \ln |f_{5}^{\ast }|+f_{5}-f_{4}\right) ^{^{\prime }},\qquad {%
h_{5}^{\ast }\neq 0;}  \label{alpha3} \\
\beta  &=&h_{5}^{\ast \ast }-\frac{h_{5}^{\ast }}{2}\left( \frac{h_{5}^{\ast
}}{h_{5}}+\frac{h_{4}^{\ast }}{h_{4}}\right) ={h_{5}^{\ast }}\left( \ln
|f_{5}^{\ast }|+f_{5}-f_{4}\right) ^{\ast },\qquad {h_{5}^{\ast }\neq 0}
\label{beta} \\
\gamma  &=&\frac{3}{2}\frac{h_{5}}{h_{5}}^{\ast }-\frac{h_{4}}{h_{4}}^{\ast
}=\left[ 3f_{5}-2f_{4}\right] ^{\ast }.  \label{gamma}
\end{eqnarray}
The partial derivatives are denoted as $h^{\bullet }=\partial h/\partial
x^{2},f^{\prime }=\partial f/\partial x^{3}$ and $f^{\ast }=\partial
f/\partial s.$ We have given the formulas both in terms of $h_{4,5}$ and
$f_{4,5}$ since we will need this later.

Formulas (\ref{ricci1a})--(\ref{ricci1d}) were obtained with
respect to anholonomic frames for a fixed linear
connection adapted to the N--connection structure, called the
canonical distinguished connection \cite{miron}. (Miron and
Anastasiei introduced this connection on vector bundles, but it
can be used in a similar fashion on (pseudo) Riemannian spaces if
the N--connection is considered). The coefficients of a
distinguished connection $\Gamma _{\ \beta \gamma }^{\alpha
}=\left( L_{\ jk}^{i},L_{\ bk}^{a},C_{\ jc}^{i},C_{\
bc}^{a}\right) ,$ are computed from the formulas
\begin{eqnarray}
L_{\ jk}^{i} &=&\frac{1}{2}g^{in}\left( \delta _{k}g_{nj}+\delta
_{j}g_{nk}-\delta _{n}g_{jk}\right) ,  \label{dcon} \\
L_{\ bk}^{a} &=&\partial _{b}N_{k}^{a}+\frac{1}{2}h^{ac}\left(
\delta _{k}h_{bc}-h_{dc}\partial _{b}N_{k}^{d}-h_{db}\partial
_{c}N_{k}^{d}\right) ,
\nonumber \\
C_{\ jc}^{i} &=&\frac{1}{2}g^{ik}\partial _{c}g_{jk},\ C_{\
bc}^{a}=\frac{1}{ 2}h^{ad}\left( \partial _{c}h_{db}+
\partial _{b}h_{dc}-\partial _{d}h_{bc}\right) .  \nonumber
\end{eqnarray}
The coefficients in (\ref{dcon}) reduce to the Christoffel symbols
if the metric components $g_{ij}$ depend only on $x$--variables,
the $h_{ab}$ depend only on $y$--variables, and the N--connection
vanishes. We emphasize that if the anholonomic frames are
introduced into consideration, there is a certain class  of linear
connections which satisfy the metricity condition for a given
metric, or inversely, there is a certain class of metrics which
satisfy the metricity conditions for a given linear connection
(this result was originally obtained by A. Kawaguchi
\cite{kawaguchi} in 1937. Details can be found in \cite{miron};
see Theorems 5.4 and 5.5 in Chapter III). So, we need to state explicitly
what type of linear connection is used for the definition of the curvature
and Ricci tensor if the space--time is provided with an anholonomic frame
structure. In this work and in Refs. \cite{vsbd,v} the linear
connection is considered to be of the form (\ref{dcon}).
The off--diagonal metrics studied in this paper
will be compatible with the canonical linear connection, but may
not have a trivial limit to a diagonal holonomic metric.

The scalar curvature is
\[
R=2\left( R_2^2+R_4^4\right) .
\]
using this along with the components of the Ricci tensor in Eqs. (\ref
{ricci1a})-(\ref{ricci1d}) one can show that for the metric ansatz (\ref
{ansatz0}) the coefficients of the energy--momentum d--tensor satisfy
\[
\Upsilon _1^1=\Upsilon _2^2+\Upsilon _4^4 , \qquad \Upsilon _2^2=\Upsilon
_3^3=\Upsilon _2 , \qquad \Upsilon _4^4=\Upsilon _5^5=\Upsilon _4,
\]
with respect to anholonomic bases (\ref{dder}) and (\ref{ddif}). Thus the
Einstein equations can be written as
\begin{equation}  \label{einsteq2a}
R_2^2 =-\kappa \Upsilon _4, \qquad R_4^4 =-\kappa \Upsilon _2, \qquad
R_{4\widehat{i}} =\kappa \Upsilon _{4\widehat{i}}, \qquad R_{5\widehat{i}}
=\kappa \Upsilon _{5\widehat{i}}, \qquad
\end{equation}
where $\widehat{i}=2,3$.

With this setup it is possible to construct very general classes of
solutions to these equations \cite{vsbd,v} which describe locally
anisotropic solitons, black holes, black tori and wormhole solutions.

\subsection{General properties of the anisotropic vacuum solutions}

In the vacuum case Eqs. (\ref{einsteq2a}) reduce to:
\begin{eqnarray}
g_3^{\bullet \bullet }-\frac{g_2^{\bullet }g_3^{\bullet }}{2g_2}-\frac{
(g_3^{\bullet })^2}{2g_3}+g_2^{^{\prime \prime }}-\frac{g_2^{^{\prime
}}g_3^{^{\prime }}}{2g_3}-\frac{(g_2^{^{\prime }})^2}{2g_2} &=&0,
\label{einsteq3a} \\
h_5^{**}-\frac{h_5^{*}}{2}\left( \frac{h_5^{*} }{h_5}+\frac{h_4^{*}}{h_4}
\right) &=&0,  \label{einsteq3b} \\
\beta w_{2,3}+\alpha _{2,3} &=&0,  \label{einsteq3c} \\
n_{2,3}^{**}+\gamma n_{2,3}^{*} &=&0.  \label{einsteq3d}
\end{eqnarray}

We now discuss general features for the d--metric coefficients, $\left(
g_2,g_3\right) ,\left( h_4,h_5\right) ,$ and the N--connection coefficients
$w_{2,3}$ and $n_{2,3}$ which solve this system of equations:

\begin{enumerate}
\item  Eq. (\ref{einsteq3a}) relates two functions $g_2(x^2,x^3)$ and $
g_3(x^2,x^3)$ and their partial derivatives in the isotropic coordinates $x^2
$ and $x^3.$ If one of the functions is fixed, by some symmetry and boundary
conditions the second function is found by solving a second order partial
differential equation. For example by redefinition of the coordinates or a
conformal transformation one can transform $g_3,$ (or conversely, $g_2$)
into a constant. Using this technique one of the authors (SV) was able to
construct various 2D soliton -- dilaton and black hole-like configurations
\cite{v}

\item  Eq. (\ref{einsteq3b}) contains partial derivatives of only the
anisotropic coordinate $s$, and relates the two functions $h_4(x^2,x^3,s)$
and $h_5(x^2,x^3,s)$. By fixing one of these functions the second one is
found by solving a second or first order differential equation in $s$ (the $x
$--variables being treated as parameters). These equations reduce to the
Bernoulli equations \cite{kamke}, and are satisfied by two arbitrary
functions $h_{4,5}(x^2,x^3)$ for which $h_{4,5}^{*}=0.$ Thus there are three
classes of solutions:

\begin{itemize}
\item  Class A, for which $h_4^{*}=0,$ $h_5^{*}\neq 0;$

\item  Class B, for which $h_5^{*}=0,$ $h_4^{*}\neq 0;$

\item  Class C, both $h_{4,5}^{*}\neq 0.$
\end{itemize}

If the condition $h_5^{*}\neq 0$ is satisfied, we can write (\ref{einsteq3b}),
in $f$--variables (see (\ref{beta})), as
\[
\left( \ln |f_5^{*}|+f_5-f_4\right) ^{*}=0,
\]
which is solved by arbitrary functions $f_5(x^2,x^3,s)$ and
\begin{equation}
f_4 (x^1,x^2,s)=f_{4[0]} (x^1,x^2) +\ln |f_5^{*}|+f_5,  \label{sole3b}
\end{equation}
Bracketed subscripts indicate ``constants'' of integration with respect to
the $s$ variable. The general solution of (\ref{einsteq3b}) expressing $h_5$
via $h_4$ is
\begin{eqnarray}
h_5(x^2,x^3,s) &=&\left[ h_{5[1]}(x^2,x^3)+h_{5[2]}(x^2,x^3)\int \sqrt{%
h_4(x^2,x^3,s)}ds,\right] ^2  \nonumber \\
&=& h_{5[0]}(x^2,x^3)[1+\varpi (x^2,x^3)s]^2, \qquad h_4^{*}=0,
\label{zeroht}
\end{eqnarray}
the integration ``constants'', $f_{5[0,1,2]}(x^2,x^3)$ and $\varpi (x^2,x^3)$
, are determined by boundary conditions and locally anisotropic limits as
well as from the requirement that the Eqs. (\ref{einsteq3c}) and (\ref
{einsteq3d}) are compatible. Conversely, for a given $h_5,$ the general
solution of (\ref{einsteq3b}) is (\ref{sole3b}) which can be rewritten with
respect to variables $h_{4,5},$ as
\begin{equation}
h_4 (x^2,x^3,s) =h_{4[0]}(x^2,x^3)\left[ \left( \sqrt{|h_5(x^2,x^3,s)|}
\right)^{*}\right] ^2  \label{sole3b1}
\end{equation}

\item  If the functions $h_{4}(x^{2},x^{3},s)$ and $h_{5}(x^{2},x^{3},s)$
are known, then Eqs. (\ref{einsteq3c}) become linearly independent algebraic
equations for $w_{2,3}$
\[
w_{2,3}\beta +\alpha _{2,3}=0,
\]
If in the case of vacuum Einstein equations $h_{5}^{\ast }=0,$ we have
$\alpha _{\widehat{i}}=\beta =0$ $\ $\ and as a consequence, Eq. (\ref
{einsteq3c}) becomes trivial, allowing arbitrary values of the functions
$w_{2,3}\left( x^{2},x^{3},s\right) $. For $h_{5}^{\ast }\neq 0$ we must
impose the condition $\alpha _{2,3}=0,$ or identify these values with the
corresponding non--diagonal components of the energy--momentum tensor. We
also note that ansatz (\ref{ansatz0}) admits an arbitrary function
$w_{1}(x^{2})$ which is not contained in the vacuum Einstein equations. This
function can be fixed by requiring that it be compatible with some locally
isotropic solutions.

\item  Eqs. (\ref{einsteq3d}) can be solved in general form if the functions
$h_{4}(x^{2},x^{3},s)$ and $h_{5}(x^{2},x^{3},s)$ (and therefore the
coefficient $\gamma $ from (\ref{gamma}) ) are known,
\begin{eqnarray}
n_{2,3}(x^{2},x^{3},s)
&=&n_{2,3[0]}(x^{2},x^{3})+n_{2,3[1]}(x^{2},x^{3})\int \frac{
h_{4}(x^{2},x^{3},s)}{{h}_{5}^{3/2}(x^{2},x^{3},s)}ds,\qquad \gamma \neq 0;
\label{ncoef} \\
n_{2,3}(x^{2},x^{3},s)
&=&n_{2,3[0]}(x^{2},x^{3})+n_{2,3[1]}(x^{2},x^{3})s,\qquad \gamma =0,
\nonumber
\end{eqnarray}
where the functions $n_{2,3[0]}(x^{2},x^{3})$ and $n_{2,3[1]}(x^{2},x^{3})$
are defined from some boundary conditions. Again the ansatz (\ref{ansatz0})
admits another arbitrary function $n_{1}(x^{2})$ which is not contained in
the vacuum Einstein equations. This function can be fixed by requiring
compatibility with some locally isotropic solutions.
\end{enumerate}

If the metric coefficients $h_{4}$ and $h_{5}$ are solutions to Eq. (\ref
{einsteq3b}) then one can define two new $[\widehat{h}_{4}=\eta _{4}h_{4},
\; \widehat{h}_{5}=\eta _{5}h_{5}]$ solutions. We call the functions $\eta
_{4,5}=\eta _{4,5}(x^{2},x^{3},s)$ gravitational polarizations since they
modify the behavior of the metric coefficients $h_{4}$ and $h_{5}$ in a
manner similar to how a material modifies the behavior of electric and
magnetic fields in media.

The  ``renormalization'' of $h_{4,5}$ into $\widehat{h}_{4,5}$ results in
the ``renormalization'' of $n_{2,3}$: $n_{2,3}\rightarrow \widehat{n}_{2,3}$
from formula (\ref{ncoef}) with $h_{4,5}\rightarrow \widehat{h}_{4,5}$ and
$\gamma \rightarrow \widehat{\gamma }.$

\section{Locally Isotropic Wormholes, Flux Tubes, and Anisotropic Running of
Constants}

We give a brief review of the locally isotropic wormhole and flux tube
solutions (DS-solutions) constructed in Refs. \cite{dzhsin,ds}, and their
anisotropic generalization proposed in Ref. \cite{vsbd}. The isotropic
DS-solutions represent 5D gravitational field configurations
which carry ``electric'' and/or ``magnetic'' charges. Various authors
have studied related 5D solutions: Liu and Wesson investigated 5D
solitonic solutions \cite{liu}; they also considered 5D charged black holes
\cite{liu2}; 5D wormhole configurations with electromagnetic charges
were studied in Ref. \cite{agnese}; Ref. \cite{ponce} looks at
5D solutions with magnetic charge; a general reference for 5D
Kaluza-Klein theory and solutions is \cite{over}. The anisotropic
constructions considered in this paper are slightly different from Ref.
\cite{vsbd}. Here we use a fixed conformal factor, in order to construct
anisotropic solutions with background geometries more general than
spherical. The study of these more general background geometries will be
carried out in section ${\bf V}$.

\subsection{5D Locally isotropic wormholes and flux tubes}

Ref. \cite{ds} considered the following spherically symmetric 5D metric,
with off--diagonal terms
\begin{eqnarray}
ds_{(DS)}^{2} &=&e^{2\nu (r)}dt^{2}-dr^{2}-a(r)(d\theta ^{2}+\sin ^{2}\theta
d\varphi ^{2})  \label{ansatz1a} \\
&-&r_{0}^{2}e^{2\psi (r)-2\nu (r)}\left[ d\chi _{(DS)}+\omega (r)dt+n\cos
\theta d\varphi \right] ^{2},  \nonumber
\end{eqnarray}
$\chi _{(DS)}$ is the 5$^{th}$ coordinate; $r,\theta ,\varphi $ are $3D$
spherical coordinates; $n$ is an integer; $r\in \{-R_{0},+R_{0}\}$ ($%
R_{0}\leq $ $\infty $) and $r_{0}$ is a constant. All functions $\nu
(r),\psi (r)$ and $a(r)$ were considered to be even functions of $r$
satisfying $\nu ^{\prime }(0)=\psi ^{\prime }(0)=a^{\prime }(0)=0$. Here we
shall study a particular class of this metric, with $\nu (r)=0$. We also
introduce a new 5$^{th}$ coordinate
\[
\chi =\chi _{(DS)}-\mu (\theta ,\varphi )^{-1}\int d\xi (\theta ,\varphi )
\]
for which
\[
d\chi _{(DS)}+n\cos \theta d\varphi =d\chi +n\cos \theta d\theta
\]
and
\[
\frac{\partial \xi }{\partial \varphi }=\mu n\cos \theta ,\qquad \frac{%
\partial \xi }{\partial \theta }=-\mu n\cos \theta ,
\]
if the factor $\mu (\theta ,\varphi )$ is taken, for instance,
\[
\mu (\theta ,\varphi )=\exp (\theta -\varphi )\  |\cos \theta |^{-1}.
\]
This redefinition of the 5$^{th}$ coordinate $\chi _{(DS)}\rightarrow \chi $%
, with $d\chi $ elongated by N--coefficients proportional to $t,r,\theta $
(isotropic coordinates) allows us to consider anisotropies on coordinates $%
\left( \varphi ,\chi \right) $. The metric (\ref{ansatz1a}), in coordinates $%
(t,r,\theta ,\varphi ,\chi ),$ and for $\nu (r)=0$ is equivalently rewritten
as
\begin{equation}
ds_{(DS)}^{2}=dt^{2}-dr^{2}-a(r)(d\theta ^{2}+\sin ^{2}\theta d\varphi
^{2})-r_{0}^{2}e^{2\psi (r)}\left[ d\chi +\omega (r)dt+n\cos \theta d\theta %
\right] ^{2}  \label{ansatz1}
\end{equation}
This form of the metric will be used to find new, anisotropic solutions of
Einstein's equations. The coefficient $\omega (r)$ in (\ref{ansatz1}) is
treated as the $t$ --component of the electromagnetic potential and $n\cos
\theta $ as the $\theta $-component. These electromagnetic potentials lead
to the metric having radial Kaluza-Klein ``electrical'' and ``magnetic''
fields. The 5D Kaluza-Klein ``electric'' field is
\begin{equation}
E_{KK}=r_{0}\omega ^{\prime }e^{3\psi }=q_{0}/a(r)  \label{emf}
\end{equation}
the ``electric'' charge $q_{0}=r_{0}\omega ^{\prime }(0)$ can be
parametrized as
\[
q_{0}=2\sqrt{a(0)}\sin \alpha _{0}.
\]
The corresponding dual, ``magnetic'' field is
\begin{equation}
H_{KK}=Q_{0}/a(r)  \label{hmf}
\end{equation}
with ``magnetic'' charge $Q_{0}=nr_{0}$ parametrized as
\[
Q_{0}=2\sqrt{a(0)}\cos \alpha _{0},
\]
The following ``circle'' relation
\begin{equation}
\frac{(q_{0}^{2}+Q_{0}^{2})}{4a(0)}{=1}  \label{emfm}
\end{equation}
relates the ``electric'' and ``magnetic'' charges. As the free parameters of
the metric are varied there are five classes of solutions with the
properties:

\begin{enumerate}
\item  $Q_0=0$ or $H_{KK}=0$, a wormhole--like ``electric'' object;

\item  $q_0=0$ or $E_{KK}=0$, a finite ``magnetic'' flux tube;

\item  $q_0=Q_0$ or $H_{KK}=E_{KK}$, an infinite ``electromagnetic'' flux
tube;

\item  $Q_0< q_0$ or $H_{KK}<E_{KK}$, a wormhole--like ``electromagnetic''
object;

\item  $Q_0> q_0$ or $H_{KK}>E_{KK}$, a finite, ``magnetic--electric'' flux
tube.
\end{enumerate}

Metric (\ref{ansatz1}) is a particular example of a d--metric of type (\ref
{dmetric}), with the ansatz functions given by (\ref{ansatz0}), or
equivalently (\ref{ansatzd}). For the coordinates $x^1 =t , x^2=r ,
x^3=\theta , y^4 = s=\chi , y^5=p=\varphi$ the set of ansatz functions
\begin{eqnarray}  \label{set1}
g_1 &=&1 , \qquad g_2=-1 , \qquad g_3=-a(r),  \nonumber \\
h_4 &=&-a(r)\sin ^2\theta ,\qquad h_5 =-r_0^2e^{2\psi (r)}, \\
w_i &=&0 ,\qquad n_1=\omega (r) , \qquad n_2=0 , \qquad n_3=n\cos \theta
\nonumber
\end{eqnarray}
defines a trivial, locally isotropic solution of the vacuum Einstein
equations (\ref{einsteq3a})--(\ref{einsteq3d}) which satisfies the
conditions $h_{4,5}^{*}=0.$ We next deal with anisotropic deformations of
this solution.

\subsection{Anisotropic generalizations of DS--solution}

The simplest way to obtain anisotropic wormhole / flux tube solutions \cite
{vsbd} is to take $r_0^2$ from (\ref{ansatz1}) or (\ref{set1})) not as a
constant, but as ``renormalized'' via $\,r_0^2\to \widehat{r}_0^2=\widehat{r}
_0^2(r,\theta ,s).$

\subsubsection{DS-solutions with anisotropy via $s=\protect\chi$}

From the isotropic solution (\ref{set1}) we generate an anisotropic solution
of Class A by taking
\[
\widehat{h}_{4}(r,\theta )=h_{4}(r,\theta )=-a(r)\sin ^{2}\theta ,
\]
with $\eta _{4}=1$ so that $\widehat{h}_{4}^{\ast }=h_{4}^{\ast }=0,$ but $%
\widehat{h}_{5}^{\ast }=\eta _{5}^{\ast }(r,\theta ,\chi )h_{5}(r)\neq 0.$
Using Eq. (\ref{zeroht}) we parametrize
\begin{equation}
\widehat{r}_{0}^{2}(\chi )\simeq r_{0(0)}^{2}[1+\varpi (r,\theta )\chi ]^{2}
\label{slin}
\end{equation}
so that
\begin{eqnarray*}
h_{5}(r,\theta ,\chi ) &=&\eta _{5}(r,\theta ,\chi )h_{5}(r) \\
h_{5[0]}(r,\theta ) &=&h_{5}(r)=-r_{0}^{2}e^{2\psi (r)},\qquad \eta _{5}(r,\theta
,\chi )=[1+\varpi (r,\theta )\chi ]^{2}.
\end{eqnarray*}
Under the conditions in this subsection $\beta $ and $\alpha _{2,3}$ from
Eqs. (\ref{alpha2}) (\ref{alpha3}) (\ref{beta}) (and therefore $w_{2,3}$)
can be arbitrary functions. Here we will require $w_{2,3}\rightarrow 0$ in
the locally isotropic limit, $\varpi \chi \rightarrow 0$. From Eq. (\ref
{ncoef}) $n_{2,3}$ depends on the anisotropic variable $s=\chi $ in the
following way
\[
n_{3}(r,\theta ,\chi )=n_{3[0]}(r,\theta )+n_{3[1]}(r,\theta )[1+\varpi
_{0}\chi ]^{-2}
\]
with $\varpi (r,\theta )=\varpi _{0}=const$. We obtain the locally isotropic
limit of (\ref{set1}), for $\varpi \chi \rightarrow 0$ if we fix the
boundary conditions with $n_{2[0,1]}=0,n_{3[0]}=0$, $n_{3[1]}(r,\theta
)=n\cos \theta $ and $n_{1}=\omega (r).$

The 5D gravitational vacuum polarization induced by variation of
``constant'' $\widehat{r}_{0}(\chi )$ renormalizes the electromagnetic
charge as $q(\chi )=\widehat{r}_{0}(\chi )\omega ^{\prime }(r=0)$. In terms
of the angular parametrization the ``electric'' charge becomes
\[
q(\chi )=2\sqrt{a(0)}\sin \alpha (\chi ),
\]
The ``electric'' field from (\ref{emf}) becomes
\[
E_{KK}=\frac{q(\chi )}{a(r)}.
\]
The renormalization of the magnetic charge, $Q_{0}\rightarrow Q(\chi )$, can
be obtained using the renormalized ``electric'' charge in relationship (\ref
{emfm}) and solving for $Q(\chi ).$ The form of (\ref{emfm}) implies that
the running of $Q(\chi )$ will be the opposite that of $q(\chi )$. For
example, if $q(\chi )$ increases with $\chi $ then $Q(\chi )$ will decrease.
The locally anisotropic polarizations $\alpha (\chi )$ are either defined
from experimental data or computed from a quantum model of 5D gravity. With
the coordinates taken as $x^{1}=t,x^{2}=r,x^{3}=\theta ,y^{4}=s=\chi
,y^{5}=p=\varphi $, one can construct a locally anisotropic solution of the
vacuum Einstein equations (\ref{einsteq3a})--(\ref{einsteq3d}) by making the
following identifications for the ansatz functions from (\ref{ansatz1})
\begin{eqnarray}
g_{1} &=&1,\qquad g_{2}=-1,\qquad g_{3}=-a(r),  \label{set2} \\
\widehat{h}_{4} &=&h_{4}=-a(r)\sin ^{2}\theta ,\qquad \eta _{4}=1  \nonumber
\\
\widehat{h}_{5} &=&\eta _{5}h_{5},\qquad h_{5}(r)=-r_{0}^{2}e^{2\psi
(r)},\qquad \eta _{5}=[1+\varpi _{0}\chi ]^{2}  \nonumber \\
w_{i} &=&0,\qquad n_{1}=\omega (r),\qquad n_{2}=0,\qquad n_{3}=n\cos \theta
\lbrack 1+\varpi _{0}\chi ]^{-2},  \nonumber
\end{eqnarray}
This generalizes the DS--solution (\ref{ansatz1}) by allowing the
Kaluza--Klein electric and magnetic charges to be dependent on ({\it i.e.}
scale with) the 5$^{th}$ coordinate $s=\chi $. We will call these the $\chi$
--solutions).

\subsubsection{DS-solutions with anisotropy via $\protect\varphi $}

In a similar fashion we can consider anisotropic dependencies with respect
to $s=\varphi $. These will be called $\varphi $--solutions. The simplest
option is to take $h_{5}^{\ast }=0$ but $h_{4}^{\ast }\neq 0,$ {\it i.e.} to
define a solution with
\begin{eqnarray*}
{\widehat{h}}_{4}(r,\theta ,\varphi ) &=&\eta _{4}h_{4}(r),\qquad
h_{4}(r)=-r_{0}^{2}e^{2\psi (r)} \\
\widehat{h}_{5}(r,\theta ) &=&h_{5}(r,\theta )=-a(r)\sin ^{2}\theta , \\
\eta _{4} &=&\exp [\varpi (r,\theta ,\varphi )],\qquad \eta _{5}=1,
\end{eqnarray*}
this allows $w_{2,3}$ to take arbitrary values since $\beta $ and $\alpha
_{2,3}$ from Eqs. (\ref{einsteq3c}) vanish. For small polarizations we can
approximate
\[
{\widehat{h}}_{4}(r,\theta ,\varphi )=h_{4[0]}(r,\theta )[1+\varpi \varphi ].
\]
The general solution of (\ref{einsteq3d}) for ${\widehat{\gamma }}=-(\ln |{%
\widehat{h}_{4}}|)^{\ast }$ is
\[
n_{2,3}(r,\theta ,\varphi )=n_{2,3[0]}(r,\theta )+n_{2,3[1]}(r,\theta )\int
\exp \varpi (r,\theta ,\varphi )d\varphi ;
\]
we take $w_{1}=\omega (r),w_{2}=0,w_{3}=n\cos \theta ,$ $n_{2[0]}(r,\theta
)=0$ and $n_{3[0]}(r,\theta )=n\cos \theta ,$ $n_{,3[1]}(r,\theta )=1$ which
are compatible with the local isotropic limit ({\it i.e.} $\varpi (r,\theta
,\varphi )\rightarrow 0$ and $\int \varpi (r,\theta ,\varphi )d\varphi
\rightarrow 0$). Taking the coordinates as $x^{1}=t,x^{2}=r,x^{3}=\theta
,y^{4}=s=\varphi ,y^{5}=p=\chi $ the following form for the ansatz functions
\begin{eqnarray}
g_{1} &=&1,\qquad g_{2}=-1,\qquad g_{3}=-a(r),  \label{set2b} \\
\widehat{h}_{4} &=&\eta _{4}h_{4},\qquad h_{4}=-r_{0}^{2}e^{2\psi
(r)},\qquad \eta _{4}=\exp [\varpi (r,\theta ,\varphi )],  \nonumber \\
\widehat{h}_{5} &=&h_{5},\qquad h_{5}=-a(r)\sin ^{2}\theta ,\qquad \eta
_{5}=1,\qquad   \nonumber \\
w_{1} &=&\omega (r),\qquad w_{2}=0,\qquad w_{3}=n\cos \theta ,  \nonumber \\
n_{1} &=&0,\qquad n_{2,3}=n_{2,3[1]}(r,\theta )\int \exp \varpi (r,\theta
,\varphi )d\varphi ,  \nonumber
\end{eqnarray}
gives a locally anisotropic generalization of the DS--metric (\ref{ansatz1})
for anisotropic dependencies on the angle $\varphi $.

We have constructed two classes of locally anisotropic generalizations of
the DS--solution : $s=\varphi$ ({\it i.e.} anisotropic angular
polarizations) or $s=\chi $ ({\it i.e.} dependence of the Kaluza--Klein
charges on the 5$^{th}$ coordinate). If the metric (\ref{ansatz1}),
describing these two classes of solutions, were given with respect to a
coordinate frame (\ref{pder}) non--diagonal terms would occur, and the study
of these solutions would be more difficult.

\section{Gravitational $\protect\theta$--Polarization of Kaluza-Klein Charges%
}

We can further generalize the forms (\ref{set2}) and (\ref{set2b}) to
generate new solutions of the 5D vacuum Einstein equations with deformations
of the constants $r_0^2$ and $n$ with respect to the $\theta$ variable.
These $\theta$ deformations take the form of the equation for an ellipsoid
in polar coordinates. This again leads to varying electric, $q,$ and
magnetic, $Q,$ charges.

\subsection{Gravitational renormalization of Kaluza-Klein charges via
variable $r_0$}

In this subsection we give a solution for which the Kaluza-Klein charges are
gravitationally renormalized by the radius becoming dependent on $\theta$ (%
{\it i.e.} in Eq. (\ref{emfm}) $a(0) \rightarrow a(\theta )$.

\subsubsection{$\protect\theta$--renormalization of charges for $\protect\chi
$--solutions}

The easiest way to obtain such $\theta $--polarizations for the $\chi $%
--solutions of (\ref{zeroht}) and (\ref{slin}) is to consider the
coordinates as $x^{1}=t,x^{2}=r,x^{3}=\theta ,y^{4}=s=\chi ,y^{5}=p=\varphi $%
, and let the ansatz functions take the form
\begin{eqnarray}
g_{1} &=&1,\qquad g_{2}=-1,\qquad g_{3}=-a(r),  \label{set4a} \\
\widehat{h}_{4} &=&h_{4}=-a(r)\sin ^{2}\theta ,\qquad \eta _{4}=1,  \nonumber
\\
\widehat{h}_{5} &=&\eta _{5}h_{5}(r),\qquad h_{5}(r)=-r_{0}^{2}e^{2\psi
(r)},\qquad \eta _{5}=\left[ 1+\varepsilon _{r}\cos \theta \right]
^{-2}[1+\varpi _{0}\chi ]^{2},  \nonumber \\
w_{i} &=&0,\qquad n_{1}=\omega (r),\qquad n_{2}=0,\qquad n_{3}=n\cos \theta
\ [1+\varpi _{0}\chi ]^{-2},  \nonumber
\end{eqnarray}
where $\eta _{5}$ is the polarizaton  and
\[
\widehat{r}_{0}^{2}(r,\theta ,\chi )\simeq r_{0(0)}^{2}\left[ 1+\varepsilon
_{r}\cos \theta \right] ^{-2}[1+\varpi _{0}\chi ]^{2},
\]
where $\varepsilon _{r}$ is the eccentricity. The ``constant'', $\widehat{r}%
_{0}$, has both an elliptic variation in $\theta $ ({\it i.e.} $r_{0(0)}%
\left[ 1+\varepsilon _{r}\cos \theta \right] ^{-1}$) and a linear variation
on the 5$^{th}$ coordinate ({\it i.e.} $[1+\varpi (r,\theta )\chi ]$).

For these $\chi $--solutions with elliptic variations as given above the 5D
Kaluza-Klein charges get renormalized through the elliptic variation of $%
\widehat{r}_0(r,\theta ,\varepsilon _r,\chi )$. This renormalizes the
``electric'' charge as
\[
q(\theta ,\chi )=r_0\sqrt{\eta _5(\theta ,\varepsilon _r,\chi )}\omega
^{\prime }(0)=\sqrt{\eta _5(\theta ,\varepsilon _r,\chi )}q_0
\]
or in terms of the angular parametrization
\[
q(\theta ,\chi )=2\sqrt{a(0)}\sqrt{\eta _5(\theta ,\varepsilon _r,\chi )}
\sin \alpha _0 ,
\]
The ``electric'' field (\ref{emf}) transforms into
\[
E_{KK}=\frac{q(\theta ,\chi )}{a(r)}=\frac{q_0}{(\sqrt{\eta _5})^{-1}a(r)}
\]
$(\sqrt{\eta _5})^{-1}$ can be treated as an anisotropic, gravitationally
induced permittivity. The renormalization of the magnetic charge, $%
Q_0\rightarrow Q(\theta ,\chi )$, can be obtained from Eq. (\ref{emfm})
using $q(\theta ,\chi )$ from above. In this case the corresponding dual,
``magnetic'' field is $H_{KK}=Q(\theta ,\chi)/a(r)$ with the ``magnetic''
charge $Q_0=nr_0$ given by
\[
Q=2\sqrt{a(0)}\sqrt{\eta _5(\theta ,\varepsilon _r,\chi )}\cos \alpha _0,
\]
These gravitationally polarized charges satisfy the circumference equation (%
\ref{emfm}) with variable radius $2\sqrt{a(0)}\sqrt{\eta_5(\theta
,\varepsilon _r,\chi )}$
\begin{equation}
\frac{(q_0^2+Q_0^2)}{4a(0)\eta _5(\theta ,\varepsilon _r,\chi )}{=1.}
\label{emfm1}
\end{equation}

\subsubsection{Elliptic renormalization of charges for $\protect\varphi$%
--solutions}

The $\varphi $--solutions can also be modified to have an elliptic variation
with respect to $\theta $. As in the case of the $\chi $--solutions this
gives an effective gravitational renormalization of the charges. With the
coordinates defined as $x^{1}=t,x^{2}=r,x^{3}=\theta ,y^{4}=s=\varphi
,y^{5}=p=\chi $ the form of this variation of the $\varphi $ -- solutions is
\begin{eqnarray}
g_{1} &=&1,\qquad g_{2}=-1,\qquad g_{3}=-a(r),  \label{set4b} \\
\widehat{h}_{4} &=&\eta _{4}h_{4},\qquad h_{4}=-r_{0}^{2}e^{2\psi
(r)},\qquad \eta _{4}=\left[ 1+\varepsilon _{r}\cos \theta \right] ^{-2}\exp
[\varpi (r,\theta ,\varphi )],  \nonumber \\
\widehat{h}_{5} &=&h_{5}=-a(r)\sin ^{2}\theta ,\qquad \eta _{5}=1,\qquad
\widehat{h}_{5}^{\ast }=h_{5}^{\ast }=0,  \nonumber \\
w_{1} &=&\omega (r),\qquad w_{2}=0,\qquad w_{3}=n\cos \theta ,  \nonumber \\
n_{1} &=&0,\qquad n_{2,3}=n_{2,3[1]}(r,\theta )\int \exp \varpi (r,\theta
,\varphi )d\varphi ,  \nonumber
\end{eqnarray}
The renormalized charges arise as in the previous example via $%
r_{0}^{2}\rightarrow \widehat{r}_{0}^{2}(r,\theta ,\varphi )$. The
``electric'' charge becomes
\[
q(\theta ,\varphi )=r_{0}\sqrt{\eta _{4}(\theta ,\varepsilon _{r},\varphi )}%
\omega ^{\prime }(0)=\sqrt{\eta _{4}(\theta ,\varepsilon _{r},\varphi )}q_{0}
\]
in terms of the angular parametrization this becomes
\[
q(\theta ,\varphi )=2\sqrt{a(0)}\sqrt{\eta _{4}(\theta ,\varepsilon
_{r},\varphi )}\sin \alpha ,
\]
The ``electric'' field (\ref{emf}) transforms into
\[
E_{KK}=\frac{q(\theta ,\varphi )}{a(r)}=\frac{q_{0}}{(\sqrt{\eta _{4}}%
)^{-1}a(r)}
\]
$(\sqrt{\eta _{4}})^{-1}$ can be treated as an anisotropic gravitationally
induced permittivity depending on the angular variables. The renormalized
magnetic charge, $Q_{0}\rightarrow Q(\theta ,\varphi )$, can be determined
using Eq. (\ref{emfm}) and $q(\theta ,\varphi )$ giving
\[
Q(\theta ,\varphi )=2\sqrt{a(0)}\sqrt{\eta _{4}(\theta ,\varepsilon
_{r},\varphi )}\cos \alpha _{0},
\]
The ``magnetic'' field is then $H_{KK}=Q(\theta ,\varphi )/a(r)$. The
gravitationally polarized charges satisfies Eq. (\ref{emfm}) with a variable
radius of $2\sqrt{a(0)}\sqrt{\eta _{4}(\theta ,\varepsilon _{r},4)}$.
\begin{equation}
\frac{(q_{0}^{2}+Q_{0}^{2})}{4a(0)\eta _{4}(\theta ,\varepsilon _{r},\varphi
)}{=1.}  \label{emfm2}
\end{equation}
There is again an elliptical variation in $\theta $, and also an anisotropic
dependence in $\varphi $.

Comparing formulas (\ref{emfm1}) and (\ref{emfm2}) we find that there are
two type of anisotropic gravitational polarizations of the charges: in the
first case the running with respect to the 5$^{th}$ coordinate is
emphasized; in the second case the anisotropy comes just from the angular
variables. In both cases there is an elliptical dependence on $\theta$.

\subsection{Gravitational renormalization of Kaluza-Klein charges via $r_0$
and $n$}

A different class of solutions from those given in Eqs. (\ref{set2}), (\ref
{set2b}) and (\ref{set4a}), (\ref{set4b}) can be constructed if, in addition
to $r_0$, we allow the $n$ in the $n \cos \theta$ term in Eq. (\ref{ansatz1}
) to vary. For the $\chi$--solutions this variable $n$ will affect $n_3$,
while for the $\varphi$--solutions it will affect $w_3$. The variability of $%
r_0$ and $n$ is parameterized using the gravitational vacuum polarizations $%
\kappa_r\left( r,\theta ,s\right)$ and $\kappa _n\left( r,\theta ,s\right)$
as
\[
r_0\rightarrow \widehat{r}_0=r_{0}/\kappa _r\left( r,\theta ,s\right) \mbox{
and }n\rightarrow \widehat{n}=n/\kappa _n\left( r,\theta ,s\right)
\]
where $\kappa_r\left( r,\theta ,s\right) =[\sqrt{\eta _4\left( r,\theta
,s\right) }]^{-1}$ or $=[\sqrt{\eta _5\left( r,\theta ,s\right) }]^{-1}$.
The polarized charges are
\[
q=q_0/\kappa _r=2\sqrt{a(0)}\sin \alpha _0/\kappa _r
\]
and
\[
Q=Q_0/\kappa _n=2\sqrt{a(0)}\cos \alpha _0/\kappa _n,
\]
Using these charges in Eq. (\ref{emfm}) gives the formula for an ellipse in
the charge space coordinates $\left(q_0,Q_0\right),$
\begin{equation}
\frac{q_0^2}{4a(0)\kappa _r^2}+\frac{Q_0^2}{4a(0)\kappa _n^2}{=1,}
\label{emfm3}
\end{equation}
the axes of the ellipse are $2\sqrt{a(0)}\kappa _r$ and $2\sqrt{a(0)} \kappa
_n.$ Formula (\ref{emfm3}) contains formulas (\ref{emfm1}) and (\ref{emfm2})
as special cases.

The form of the $\chi$ and $\varphi$ -- solutions from the previous two
subsections gets modified by this ``elliptic'' renormalization of the
Kaluza-Klein charges.

\begin{itemize}
\item  With the coordinates defined as $x^1 =t , x^2=r , x^3=\theta , y^4
=s=\chi , y^5=p=\varphi$ the $\chi $--solutions with $\kappa _r=[\sqrt{\eta
_5\left( r,\theta ,\chi \right) }]^{-1}$ take the form
\begin{eqnarray}
g_1 &=&1 , \qquad g_2=-1 , \qquad g_3=-a(r),  \nonumber \\
\widehat{h}_4 &=&h_4=-a(r)\sin ^2\theta , \qquad \widehat{h}_4^{*}=0 ,
\qquad h_4^{*}=0 , \qquad \eta_4=1,  \label{set5a} \\
\widehat{h}_5 &=&\eta _5h_5 , \qquad h_5=-r_0^2e^{2\psi (r)} , \qquad \eta
_5=1/\kappa_r^2(r,\theta ,\chi ),  \nonumber \\
w_i &=&0 , \qquad n_1=\omega (r) , \qquad n_2=0 , \qquad n_3=n\cos \theta
/\kappa _n(r,\theta ,\chi).  \nonumber
\end{eqnarray}

\item  With the coordinates defined as $x^1 =t , x^2=r , x^3=\theta , y^4
=s=\varphi , y^5=p=\chi$ the $\varphi $--solutions with $\kappa _r=[\sqrt{%
\eta _4\left( r,\theta ,\chi \right) }]^{-1}$ take the form
\begin{eqnarray}
g_1 &=&1 , \qquad g_2=-1 , \qquad g_3=-a(r),  \nonumber \\
\widehat{h}_4 &=&\eta _4h_4 , \qquad h_4=-r_0^2e^{2\psi (r)} , \qquad\eta
_4=1/\kappa_r^2(r,\theta ,\varphi ),  \label{set5b} \\
\widehat{h}_5 &=&h_5=-a(r)\sin ^2\theta , \qquad \eta _5=1 , \qquad h_5^{*}=0
\nonumber \\
w_1 &=&\omega (r) , \qquad w_2=0 , \qquad w_3=n\cos \theta /\kappa _n \left(
r,\theta,\varphi \right) ,  \nonumber \\
n_1 &=&0 , \qquad n_{2,3}=n_{2,3[1]}(r,\theta )\int \ln | \kappa
_r(r,\theta,\varphi )|d\varphi ,  \nonumber
\end{eqnarray}
\end{itemize}

\section{Wormholes in Ellipsoidal, Cylindrical, Bipolar and Toroidal
Backgrounds}

The locally anisotropic wormhole / flux tube solutions presented in the
previous sections are anisotropic deformations from a spherical 3D
hypersurface background. These solutions can be generalized to other
rotational hypersurface geometry backgrounds. In this section we will give
the explicit forms for these generalized solutions and analyze their basic
properties. The notations and metric relations for the 3D Euclidean
rotational hypersurfaces that we use will be those of Ref. \cite{korn}.

\subsection{Elongated rotation ellipsoid hypersurfaces}

An elongated rotation ellipsoid hypersurface (a 3D e--ellipsoid) is given by
the formula
\begin{equation}
\frac{x^{2}+y^{2}}{\sigma ^{2}-1}+\frac{z^{2}}{\sigma ^{2}}=\widetilde{a}%
^{2}(r),  \label{relhor}
\end{equation}
where $\sigma \geq 1,$ and $x,y,z$ here are the usual Cartesian coordinates.
$\widetilde{a}(r)$ is similar to the radius in the spherical symmetric case.
The 3D, ellipsoidal coordinate system is defined
\begin{equation}
x=\widetilde{a}\sinh u\sin v\cos s,\qquad y=\widetilde{a}\sinh u\sin v\sin
s,\qquad z=\widetilde{a}\ \cosh u\cos v,
\end{equation}
where $\sigma =\cosh u$ and $0\leq u<\infty ,\ 0\leq v\leq \pi ,\ 0\leq
s<2\pi .$ The hypersurface metric is
\begin{equation}
g_{uu}=g_{vv}=\widetilde{a}^{2}\left( \sinh ^{2}u+\sin ^{2}v\right) ,\qquad
g_{ss}=\widetilde{a}^{2}\sinh ^{2}u\sin ^{2}v.  \label{hsuf1}
\end{equation}
It will be more useful to consider a conformally transformed metric, where
the components in Eq.(\ref{hsuf1}) are multiplied by the conformal factor $%
\widetilde{a}^{-2}\left( \sinh ^{2}u+\sin ^{2}v\right) ^{-1},$ giving
\begin{eqnarray}
ds_{(3e)}^{2} &=&du^{2}+dv^{2}+g_{ss}(u,v)ds^{2}  \label{hsuf1a} \\
g_{ss}(u,v) &=&\sinh ^{2}u\sin ^{2}v/(\sinh ^{2}u+\sin ^{2}v).  \nonumber
\end{eqnarray}

\subsection{Flattened rotation ellipsoid hypersurfaces}

In a similar fashion we consider the hypersurface equation for a flattened
rotation ellipsoid (a 3D f--ellipsoid),
\begin{equation}
\frac{x^{2}+y^{2}}{1+\sigma ^{2}}+\frac{z^{2}}{\sigma ^{2}}=\widetilde{a}%
^{2}(r),  \label{relhor1}
\end{equation}
here $\sigma \geq 0$ and $\sigma =\sinh u.$ In this case the 3D coordinate
system is defined as
\begin{equation}
x=\widetilde{a}\cosh u\sin v\cos s,\qquad y=\widetilde{a}\cosh u\sin v\sin
s,\qquad z=\widetilde{a}\sinh u\cos v,
\end{equation}
where $0\leq u<\infty ,\ 0\leq v\leq \pi ,\ 0\leq s<2\pi .$ The hypersurface
metric is
\begin{equation}
g_{uu}=g_{vv}=\widetilde{a}^{2}\left( \sinh ^{2}u+\cos ^{2}v\right) \qquad
g_{\varphi \varphi }=\widetilde{a}^{2}\sinh ^{2}u\cos ^{2}v,  \nonumber
\end{equation}
Again for later convenience we consider a conformally transformed version of
this metric
\begin{eqnarray}
ds_{(3f)}^{3} &=&du^{2}+dv^{2}+g_{ss}(u,v)ds^{2},  \label{hsuf1b} \\
g_{ss}(u,v) &=&\sinh ^{2}u\cos ^{2}v/(\sinh ^{2}u+\cos ^{2}v).  \nonumber
\end{eqnarray}

\subsection{Ellipsoidal cylindrical hypersurfaces}

The formula for an ellipsoidal cylindrical hypersurface is
\begin{equation}
\frac{x^{2}}{\sigma ^{2}}+\frac{y^{2}}{\sigma ^{2}-1}=\rho ^{2},\ z=s,
\label{elcyl}
\end{equation}
where $\sigma \geq 1.$ The 3D radial coordinate is given as $\widetilde{a}%
^{2}=\rho ^{2}+s^{2}.$ The 3D coordinate system is defined
\[
x=\rho \cosh u\cos v,\qquad y=\rho \sinh u\sin v,\qquad z=s,
\]
where $\sigma =\cosh u$ and $0\leq u<\infty ,\ 0\leq v\leq \pi $. Using the
expressions for $x,y$ and Eq. (\ref{elcyl}) we can make the change $\rho
(x,y)\rightarrow \rho (u,v)$. The hypersurface metric is
\[
g_{uu}=g_{vv}=\rho ^{2}(u,v)\ \left( \sinh ^{2}u+\sin ^{2}v\right) ,\qquad
g_{ss}=1;
\]
we will again consider a conformally transformed version of this metric
\begin{eqnarray}
ds_{(3c)}^{2} &=&du^{2}+dv^{2}+g_{ss}(u,v,\rho (u,v))ds^{2},  \label{cylin1}
\\
g_{ss}(u,v) &=&1/\rho ^{2}(u,v)\ (\sinh ^{2}u+\sin ^{2}v).  \nonumber
\end{eqnarray}

\subsection{ Bipolar coordinates}

Now we consider a bipolar hypersurface given by the formula
\begin{equation}
\left( \sqrt{x^{2}+y^{2}}-\frac{\widetilde{a}(r)}{\tan \xi }\ \right)
^{2}+z^{2}=\frac{\widetilde{a}^{2}(r)}{\sin ^{2}\xi },  \label{bip}
\end{equation}
which describes a hypersurface obtained by rotating the circles
\[
\left( y-\frac{\widetilde{a}(r)}{\tan \xi }\right) ^{2}+z^{2}=\frac{%
\widetilde{a}^{2}(r)}{\sin ^{2}\xi }
\]
around the $z$ axis; because $|\tan \xi |^{-1}<|\sin \xi |^{-1},$ the
circles intersect the $z$ axis. The relationship between the Cartesian
coordinates and the bipolar coordinates is
\[
x=\frac{\widetilde{a}(r)\sin \xi \cos s}{\cosh \tau -\cos \xi },\qquad y=%
\frac{\widetilde{a}(r)\sin \xi \sin s}{\cosh \tau -\cos \xi },\qquad z=\frac{%
\widetilde{a}(r)\sinh \tau }{\cosh \tau -\cos \xi },
\]
where $-\infty <\tau <\infty ,0\leq \xi <\pi ,0\leq s<2\pi $. The
hypersurface metric is
\[
g_{\tau \tau }=g_{\xi \xi }=\frac{\widetilde{a}^{2}(r)}{\left( \cosh \tau
-\cos \xi \right) ^{2}},\qquad g_{ss}=\frac{\widetilde{a}^{2}(r)\sin ^{2}\xi
}{\left( \cosh \tau -\cos \xi \right) ^{2}},
\]
which, after multiplication by the conformal factor $\left( \cosh \tau -\cos
\sigma \right) ^{2}/\rho ^{2}$ becomes
\begin{equation}
ds_{(3b)}^{2}=d\tau ^{2}+d\xi ^{2}+g_{ss}(\xi )ds^{2},\qquad g_{ss}(\xi
)=\sin ^{2}\xi .  \label{mbipcy}
\end{equation}

\subsection{Toroidal coordinates}

Now we consider a toroidal hypersurface with nontrivial topology given by
the formula
\begin{equation}
\left( \sqrt{x^2+y^2}-\widetilde{a}(r)\ (\coth \xi )\right) ^2+z^2= \frac{%
\widetilde{a}^2(r)}{\sinh ^2\xi },  \label{torus}
\end{equation}
the relationship to the Cartesian coordinates is given by
\[
x =\frac{\widetilde{a}(r)\sinh \tau \cos s}{\cosh \tau -\cos \xi },\qquad y=%
\frac{\widetilde{a}(r)\sin \xi \sin s}{\cosh \tau -\cos \xi }, \qquad z =
\frac{\widetilde{a}(r)\sinh \xi }{\cosh \tau -\cos \xi },
\]
where $-\pi <\xi <\pi ,0\leq \tau <\infty ,0\leq s<2\pi$. The hypersurface
metric is
\[
g_{\sigma \sigma }=g_{\tau \tau }=\frac{\widetilde{a}^2(r)}{\left( \cosh
\tau -\cos \xi \right) ^2} , \qquad g_{ss}=\frac{\widetilde{a}^2(r)\sin
^2\xi } {\left( \cosh \tau -\cos \xi \right) ^2}
\]
After multiplication by the conformal factor $\left( \cosh \tau -\cos \sigma
\right) ^2/\widetilde{a}^2(r)$ this takes the same form as (\ref{mbipcy})
\begin{equation}  \label{mtor}
ds_{(3t)}^3 =d\tau ^2+d\xi ^2+g_{ss}(\xi )ds^2, \qquad g_{ss}(\xi ) =\sin
^2\xi ,
\end{equation}
Although this looks identical to the metric in (\ref{mbipcy}) the
coordinates $\left( \tau ,\xi ,s\right)$ have different meanings in each
case. This can be seen by the different ranges for the two cases.

\subsection{Anisotropic wormholes in rotation deformed hypersurface
backgrounds}

In order to construct wormholes which exhibit the various 3D geometries
cataloged above, we will associate one of the ansatz functions of the
wormhole solutions with $g_{ss}(x^2, x^3)$. For the $\chi$--solutions this
is accomplished by letting $h_4=g_{ss}(x^2,x^3)$; for the $\varphi$%
--solutions this is accomplished letting by $h_5=g_{ss}(x^2,x^3)$.

The construction of such solutions is based on the assumption that $%
g_{ss}(x^2,x^3)$ for the five non-spherical geometries listed above is to be
taken as $h_4=g_{ss}(x^2,x^3)$ (for $\chi $--solutions), or as $%
h_5=g_{ss}(x^2,x^3)$ (for $\varphi (z)$--solutions). In each case $h_{4,5}$
is multiplied by corresponding gravitational polarizations, $\eta _{4,5}$,
so as to give wormhole/flux tube configurations of the form (\ref{set2}) , (%
\ref{set4a}) , (\ref{set5a}) (for $\chi $--solutions) or configurations of
the form (\ref{set2b}), (\ref{set4b}), (\ref{set5b}) (for $\varphi (z)$%
--solutions).

\subsubsection{The $\protect\chi$--solutions}

For the five 3D geometries given above, the d--metrics (\ref{ansatzd}) for
the $\chi $--solutions from Eqs. (\ref{set2}), (\ref{set4a}), (\ref{set5a}),
have the coordinates defined as
\begin{eqnarray}
x^{k} &=&\left\{
\begin{array}{l}
(t,u,v),0\leq u<\infty ,\ 0\leq v\leq \pi ,\cosh u\geq 1,%
\mbox{ ellipsoid
(\ref{relhor})}; \\
(t,u,v),0\leq u<\infty ,\ 0\leq v\leq \pi ,\sinh u\geq 0,%
\mbox{ ellipsoid
(\ref{relhor1})}; \\
(t,u,v),0\leq u<\infty ,\ 0\leq v\leq \pi ,\cosh u\geq 1,%
\mbox{cylinder
(\ref{elcyl})}; \\
(t,\tau ,\xi ),-\infty <\tau <\infty ,0\leq \xi <\pi ,%
\mbox{bipolar
(\ref{bip})}; \\
(t,\tau ,\xi ),0\leq \tau <\infty ,-\pi <\xi <\pi ,%
\mbox{torus
(\ref{torus})};
\end{array}
\right.   \nonumber \\
y^{4} &=&s=\chi ,\qquad y^{5}=p=\left\{
\begin{array}{l}
\varphi \in \lbrack 0,2\pi ),\mbox{ ellipsoids };\mbox{bipolar };\mbox{torus}%
; \\
z\in (-\infty ,\infty ),\mbox{cylinder};
\end{array}
\right.   \nonumber
\end{eqnarray}
and the ansatz functions given as
\begin{eqnarray}
g_{1} &=&1,\qquad g_{2}=-1,\qquad g_{3}=-1,   \label{ellipschi} \\
\widehat{h}_{4} &=&\eta _{4}h_{4},\qquad h_{4}=g_{ss}(x^{2},x^{3})=\left\{
\begin{array}{l}
\frac{\sinh ^{2}u\sin ^{2}v}{\sinh ^{2}u+\sin ^{2}v},%
\mbox{ ellipsoid
(\ref{hsuf1a})}; \\
\frac{\sinh ^{2}u\cos ^{2}v}{\sinh ^{2}u+\cos ^{2}v},%
\mbox{ellipsoid
(\ref{hsuf1b})}; \\
\frac{\rho ^{-2}(u,v)}{\sinh ^{2}u+\sin ^{2}v},\mbox{cylinder (\ref{elcyl})};
\\
\sin ^{2}\xi ,\mbox{bipolar  (\ref{bip})};\mbox{torus  (\ref{torus})};
\end{array}
\right.   \nonumber \\
\eta _{4} &=&\left[ \left( \sqrt{\widehat{h}_{5}(x^{2},x^{3},\chi )}\right)
^{\ast }\right] ^{2},\mbox{see (\ref{sole3b})}, \\
\eta _{5} &=&\left\{
\begin{array}{l}
\lbrack 1+\varpi _{0}\chi ]^{2},\mbox{ see  (\ref{set2})}; \\
\left[ 1+\varepsilon _{r}\cos x^{3}\right] ^{-2}[1+\varpi _{0}\chi ]^{2},%
\mbox{
see  (\ref{set4a})}; \\
1/\kappa _{r}^{2}(x^{2},x^{3},\chi ),\mbox{ see  (\ref{set5a})};
\end{array}
\right.   \nonumber \\
\widehat{h}_{5} &=&\eta _{5}h_{5},\qquad h_{5}(x^{2},x^{3},\chi
)=-r_{0}^{2}\exp \{2\psi \lbrack r(x^{2},x^{3},\chi )]\};\qquad w_{i}=0;
\nonumber \\
r &=&\widetilde{a}^{(invers)}\left( x^{2},x^{3},\chi \right) \mbox{ from }%
\mbox{
(\ref{relhor})};\mbox{ (\ref{relhor1})};\mbox{ (\ref{elcyl})};%
\mbox{
(\ref{bip})};\mbox{(\ref{torus})};  \nonumber \\
\ n_{1} &=&\omega (x^{2}),\qquad n_{2}=0,\qquad n_{3}=n\cos x^{2}\times
\left\{
\begin{array}{l}
\lbrack 1+\varpi _{0}\chi ]^{-2},\mbox{ see  (\ref{set2})}; \\
\lbrack 1+\varpi _{0}\chi ]^{-2},\mbox{ see  (\ref{set4a})}; \\
1/\kappa _{n}(x^{2},x^{3},\chi ),\mbox{ see  (\ref{set5a})}.
\end{array}
\right.   \nonumber
\end{eqnarray}
The formulas (\ref{ellipschi}) describe wormhole / flux tube configurations
which are defined self--consistently in the various rotational hypersurface
backgrounds listed above. As in the case of the spherical background these
solutions have an anisotropic deformation with respect to the given
hypersurface backgrounds.

\subsubsection{The $\protect\varphi$--solutions}

Now we give the form of the d--metric (\ref{ansatzd}) for the $\varphi $%
--solutions embedded in the five 3D rotational hypersurfaces. The
coordinates are taken as
\begin{eqnarray}
x^{k} &=&\left\{
\begin{array}{l}
(t,u,v),0\leq u<\infty ,\ 0\leq v\leq \pi ,\cosh u\geq 1,%
\mbox{ ellipsoid
(\ref{relhor})}; \\
(t,u,v),0\leq u<\infty ,\ 0\leq v\leq \pi ,\sinh u\geq 0,%
\mbox{ ellipsoid
(\ref{relhor1})}; \\
(t,u,v),0\leq u<\infty ,\ 0\leq v\leq \pi ,\cosh u\geq 1,%
\mbox{cylinder
(\ref{elcyl})}; \\
(t,\tau ,\xi ),-\infty <\tau <\infty ,0\leq \xi <\pi ,%
\mbox{bipolar
(\ref{bip})}; \\
(t,\tau ,\xi ),0\leq \tau <\infty ,-\pi <\xi <\pi ,%
\mbox{torus
(\ref{torus})};
\end{array}
\right.   \nonumber \\
y^{4} &=&s=\left\{
\begin{array}{l}
\varphi \in \lbrack 0,2\pi ),\mbox{ ellipsoid };\mbox{bipolar };\mbox{torus};
\\
z\in (-\infty ,\infty ),\mbox{cylinder};
\end{array}
\right. y^{5}=p=\chi ,  \nonumber
\end{eqnarray}
and the ansatz functions given as
\begin{eqnarray}
g_{1} &=&1,\qquad g_{2}=-1,\qquad g_{3}=-1,  \label{ellipschi1} \\
\widehat{h}_{4} &=&\eta _{4}h_{4},\qquad h_{4}=g_{ss}(x^{2},x^{3})=\left\{
\begin{array}{l}
\frac{\sinh ^{2}u\sin ^{2}v}{\sinh ^{2}u+\sin ^{2}v},%
\mbox{ ellipsoid
(\ref{hsuf1a})}; \\
\frac{\sinh ^{2}u\cos ^{2}v}{\sinh ^{2}u+\cos ^{2}v},%
\mbox{ellipsoid
(\ref{hsuf1b})}; \\
\frac{\rho ^{-2}(u,v)}{\sinh ^{2}u+\sin ^{2}v},\mbox{cylinder (\ref{elcyl})};
\\
\sin ^{2}\xi ,\mbox{bipolar  (\ref{bip})};\mbox{torus  (\ref{torus})};
\end{array}
\right.   \nonumber \\
\eta _{4} &=&\left\{
\begin{array}{l}
\exp [\varpi (x^{2},x^{3},s],\mbox{ see  (\ref{set2b})}; \\
\left[ 1+\varepsilon _{r}\cos x^{3}\right] ^{-2}\exp [\varpi
(x^{2},x^{3},s)],\mbox{
see  (\ref{set4b})}; \\
1/\kappa _{r}^{2}(x^{2},x^{3},s),\mbox{ see  (\ref{set5b})};
\end{array}
\right.  \\
\eta _{5} &=&\left\{
\begin{array}{l}
\eta _{5[0]}\left( x^{2},x^{3}\right) +\eta _{5[1]}\left( x^{2},x^{3}\right)
\int \eta _{4}\left( x^{2},x^{3},s\right) ds,\mbox{ for }\widehat{h}%
_{4}^{\ast }\neq 0, \\
\eta _{5[0]}\left( x^{2},x^{3}\right) +\eta _{5[1]}\left( x^{2},x^{3}\right)
s;\mbox{ for }\widehat{h}_{4}^{\ast }=0;
\end{array}
\right.   \nonumber \\
\widehat{h}_{5} &=&\eta _{5}h_{5},\qquad h_{5}(x^{2},x^{3},s)=-r_{0}^{2}\exp
\{2\psi \lbrack r(x^{2},x^{3},s)]\};  \nonumber \\
r &=&\widetilde{a}^{(invers)}\left( x^{2},x^{3},s\right) \mbox{ from }%
\mbox{
(\ref{relhor})};\mbox{ (\ref{relhor1})};\mbox{ (\ref{elcyl})};%
\mbox{
(\ref{bip})};\mbox{(\ref{torus})};  \nonumber \\
w_{1} &=&\omega (x^{2}),\qquad w_{2}=0,\qquad w_{3}=n\cos x^{2}\times
\left\{
\begin{array}{l}
1,\mbox{ see  (\ref{set2b})}; \\
1,\mbox{ see  (\ref{set4b})}; \\
1/\kappa _{n}(x^{2},x^{3},s),\mbox{ see  (\ref{set5b})}.
\end{array}
\right.   \nonumber \\
\ n_{1} &=&0,\qquad n_{2,3}=n_{2,3[0]}\left( x^{2},x^{3}\right) \times
\left\{
\begin{array}{l}
\int \exp \varpi _{0}\left( x^{2},x^{3},s\right) ds,\mbox{ see
(\ref{set2b})}; \\
\int \exp \varpi _{0}\left( x^{2},x^{3},s\right) ds,\mbox{ see
(\ref{set4b})}; \\
\int |\ln \kappa _{n}(x^{2},x^{3},s)|ds,\mbox{ see  (\ref{set5b})}.
\end{array}
\right. ^{{}}  \nonumber
\end{eqnarray}

Formulas (\ref{ellipschi1}) describe a large class of wormhole / flux tube
configurations which are defined self--consistently in the various 3D
rotation hypersurface backgrounds. The deformations in this case come from
the angular coordinate $s=\varphi$ (for the ellipsoid, bipolar and toroidal
cases) or from the axial coordinate $s=z$ (for the cylindrical case). These
``deformation'' or anisotropic coordinates, $\varphi$ or $z$, are also third
coordinates about which the rotation of the hypersurfaces occurs.

\section{Conclusions}

The construction of wormhole and/or flux tube solutions in modern string
theory, extra dimensional gravity and quantum chromodynamics is of
fundamental importance in understanding these theories (especially their
non-perturbative aspects). Such solutions are difficult to find, and the
solutions which are known usually have a high degree of symmetry. In this
paper we have applied the method of anholonomic frames to construct the
general form of wormholes and flux tubes in 5D Kaluza-Klein theory. These
solutions have local anisotropy which would make their study using holonomic
frames difficult. This helps to demonstrate the usefulness of the
anholonomic frames method in studying anisotropic solutions. Most physical
situations do not possess a high degree of symmetry, and so the anholonomic
frames method provides a useful mathematical framework for studying these
less symmetric configurations.

The key result of this paper is the demonstration that off--diagonal metrics
in 5D Kaluza--Klein theory can be parametrized into forms that define new,
interesting classes of solutions of Einstein's vacuum equations. These
solutions represent wormhole and flux tube configurations which are locally
anisotropic. These anisotropic solutions reduce to previously known
spherically symmetric wormhole metrics \cite{chodos,dzhsin,ds} in the local
isotropic limit. These anisotropic solutions also extend the idea of Salam,
Strathee and Perracci \cite{sal} that including off--diagonal components in
higher dimensional metrics gives rise to gauge fields and charges. Not only
do we find ``electric'' and``magnetic'' charges for our solutions, but the
anisotropies in the 5$^{th}$ coordinate ($\chi$) and/or in the angular
coordinate ($\varphi$) give a gravitational scaling or running of these
Kaluza-Klein charges. Such a gravitational scaling of charges could provide
an experimental signature for the presence of extra dimensions ({\it i.e.}
if some charge were observed to exhibit a running which was not in agreement
with that given by 4D quantum field theory this could be evidence for a
gravitational running of the charge).

In the first part of this paper these anisotropic solutions were constructed
as deformations from a spherical background. In the final section of this
paper we showed that it is possible to construct a large variety of such
anisotropic solutions as deformations from various background geometries:
elliptic (elongated and flattened), cylindrical, toroidal an bipolar.

\section*{Acknowledgments}

D.S. would like to thank V. Dzhunushaliev for discussions related to this
work. S. V. emphasizes the collaboration with E.\ Gaburov and D. Gontsa on
the definition of the final formulas for Einstein's equations in mixed
holonomic--anholonomic variables. The work is supported by a 2000--2001
California State University Legislative Award and
a NATO/Portugal fellowship grant at the Technical University of Lisbon.

\end{document}